\documentclass{article}

\input{tcilatex}

\begin{document}

\title{Robertson-Walker fluid sources endowed with rotation characterised by
quadratic terms in angular velocity parameter.}
\author{RJ Wiltshire \\
The Division of Mathematics \& Statistics,\\
The University of Glamorgan,\\
Pontypridd CF37 1DL, UK\\
email: rjwiltsh@glam.ac.uk}
\maketitle

\begin{abstract}
Einstein's equations for a Robertson-Walker fluid source endowed with
rotation are presented upto and including quadratic terms in angular
velocity parameter. A family of analytic solutions are obtained for the case
in which the source angular velocity is purely time-dependent. A subclass of
solutions is presented which merge smoothly to homogeneous rotating and
non-rotating central sources. The particular solution for dust endowed with
rotation is presented. In all cases explicit expressions, depending
sinusoidally on polar angle, are given for the density and internal
supporting pressure of the rotating source. In addition to the non-zero
axial velocity of the fluid particles it is shown that there is also a
radial component of velocity which vanishes only at the poles. The velocity
four-vector has a zero component between poles.
\end{abstract}

\bigskip

\section{Introduction}

Perturbation techniques, so important in the General Theory of Relativity,
have frequently been applied successfully in the description of slowly
rotating compact perfect fluid sources. Of considerable importance has been
the analysis of Hartle \cite{har} who presented the equations for the
equilibrium configurations of cold stars up to and including the first order
of angular \ velocity parameter. This work has formed the basis of an
extended first order analysis \ for example, Kojima \cite{koj1} and \cite%
{koj2} Beyer \& Kokkotas \cite{beyer}, Abramowicz \textit{et al } \cite{abro}%
, to the address the problem of the non-radial quasi-periodic oscillations
of rotating compact sources and the resulting r-mode spectrum \ of
relativistic stars .

In examples of non-compact rotating sources Kegeles \cite{keg} and Wiltshire 
\cite{Wilt} successfully applied the perturbation method to a slowly
rotating non-equilibrium configuration. In this case a Robertson-Walker dust
source endowed with rotation was successfully matched to the Kerr exterior
solution of Einstein's equations to the first order in angular velocity
parameter.

Recent second order perturbation analyses have largely been confined to
non-rotating cases. For example, Salopek \textit{et al }\cite{salo}, Russ 
\textit{et al} \cite{russ} use the method to discuss the gravitational
instabilities of an expanding inhomogeneous universe. However, there is
seemingly an absence of literature on the use of second order techniques to
describe rotating bodies. This is perhaps surprising, since such analyses
can reveal the relativistic effects of the spatial distribution of a
rotating fluid as characterised by internal density and pressure. Such
effects are not revealed in first order approximations where fluid density
and pressure are shown to be same as for a non-rotating source. Moreover the
second order effects result in calculated deviations of spherical symmetry \
in the fluid boundary of a compact or extended body which may be used in the
context of the matching problem with a Kerr vacuum.

It is the aim here to present an example of a second order perturbation
analysis applied to an extended rotating source. In particular a
Robertson-Walker source will be endowed with rotation. As will be seen the
approach naturally brings about solutions of Einstein's equations which
exhibit the non-homogeneities in internal density and supporting pressure
due to rotation.

In the following a non-rotating source will be described using the
Robertson-Walker metric in the form:

\begin{equation}
d\sigma _{RW}^{2}=d\eta ^{2}-R^{2}\left( \eta \right) \left( \frac{d\xi ^{2}%
}{1-k\xi ^{2}}+\xi ^{2}d\theta ^{2}+\xi ^{2}\sin ^{2}\left( \theta \right)
d\phi ^{2}\right)  \label{ebo5}
\end{equation}%
where $R=R\left( \eta \right) $ , $k=-1,0,1$ and where the speed of light $%
c\equiv 1$, and gravitational constant $G\equiv 1.$ The homogeneous density
and supporting pressure will be denoted by $\rho _{RW}$ and $p_{RW}$
respectively. The fluid source will be endowed with rotation which will be
characterised in terms of an angular speed parameter, denoted by $q$, and
the mathematical analysis to follow will be accurate up to and including
second order terms in $q$.

The rotating fluid source will be described in terms of an extended form of
the Robertson-Walker metric which is taken to be:

\begin{eqnarray}
d\sigma ^{2} &=&\left( 1+\tilde{Q}q^{2}\right) d\eta ^{2}-\frac{R^{2}\left(
\eta \right) }{1-k\xi ^{2}}\left( 1+\tilde{U}q^{2}\right) d\xi ^{2}-2\tilde{J%
}\xi ^{2}R^{2}q^{2}d\xi d\eta  \nonumber \\
&&-\xi ^{2}R^{2}\left( 1+\tilde{V}q^{2}\right) d\theta ^{2}-\xi
^{2}R^{2}\sin ^{2}\left( \theta \right) \left( 1+\tilde{W}q^{2}\right) d\phi
^{2}  \nonumber \\
&&-2\xi ^{2}\sin ^{2}\left( \theta \right) R^{2}q\left( Yd\xi d\phi +Xd\phi
d\eta \right) +O\left( q^{3}\right)  \label{ebo19s}
\end{eqnarray}%
in which each of the functions, $\tilde{J}$, $\tilde{Q}$, $\tilde{U}$, $%
\tilde{V}$, $\tilde{W}$ depend on $\xi $, $\theta $ and $\eta $ whilst $X$, $%
Y$ depend on $\xi $ and $\eta $ alone. Note the this form of metric is an
extension of the linearly perturbed form of the Robertson-Walker \ metric
used by Kegeles \cite{keg}. The components of the fundamental tensor from (%
\ref{ebo19s}) are determined using Einstein's equations for a perfect fluid
written here in the form: 
\begin{equation}
G_{b}^{a}=-8\pi T_{b}^{a}\quad \text{,}\quad \quad T_{b}^{a}=\left( \rho
+p\right) u^{a}u_{b}-\delta _{b}^{a}p\quad \text{,}  \label{ebo60p}
\end{equation}%
where $\rho $, $p$ are the respective rotating source density and supporting
internal pressure and $u^{a}$ are the components of the velocity four-vector
with the property that $u^{a}u_{a}=1.$

The use of seven as yet unknown functions in (\ref{ebo19s}) will naturally
gives rise to ambiguity in the solution of Einstein's equations. This
ambiguity can be removed by further specification of the gauge in which the
solutions are to be determined. Although convenient gauge choices are much
discussed in the \ literature, including recently, Bruni \textit{et al }\cite%
{bruni} it is convenient at this stage to continue the solution process
without further specification of the gauge other than that which is explicit
from the metric choice (\ref{ebo19s}).

For brevity the term $O\left( q^{3}\right) $,included in (\ref{ebo19s}) will
be omitted from all future expressions. The fact that all terms including $%
q^{n}$, $n\geq 3$ are taken as negligibly small will of course be implied.

In summary it is the aim in the following to determine solutions of
Einstein's equations in the form (\ref{ebo60p}) for rotating sources
described by the metric (\ref{ebo19s})

\section{Solution Approach}

Direct calculation of the components of the Einstein tensor for (\ref{ebo19s}%
) show that the components \ $G_{3}^{2}$ and $G_{2}^{3}$ are identically
zero and so for a rotating perfect fluid it follows that the velocity four
vector component $u^{2}=0$ and one form component $u_{2}=0$. It thus follows
that the conditions $T_{1}^{2}=0=T_{2}^{1}$ and $T_{4}^{2}=0=T_{2}^{4}$ must
hold upto and including terms in \ $q^{2}$ . Moreover $T_{2}^{2}+p=0$.

In addition the perfect fluid (\ref{ebo60p}) must also satisfy the following
consistency relationships%
\begin{eqnarray}
\left( T_{1}^{1}+p\right) \left( T_{3}^{3}+p\right) -T_{3}^{1}T_{1}^{3} &=&0
\nonumber \\
\left( T_{1}^{1}+p\right) \left( T_{4}^{4}+p\right) -T_{4}^{1}T_{1}^{4} &=&0
\nonumber \\
\left( T_{3}^{3}+p\right) \left( T_{4}^{4}+p\right) -T_{4}^{3}T_{3}^{4} &=&0
\label{ebo20}
\end{eqnarray}%
However by direct calculation $G_{1}^{3}$, $G_{3}^{1}$ , $G_{4}^{3}$ and $%
G_{3}^{4}$ depend on linear terms in $q$, whilst $G_{2}^{1}$, $G_{1}^{2}$, $%
G_{1}^{4}$, $G_{4}^{1}$, $G_{4}^{2}$ and $G_{2}^{4}$ depend on quadratic
terms in $q$ . Also when \ $q=0$ each of these components is zero. It
follows that for the Robertson-Walker source endowed with rotation that
solutions of Einstein's equations must satisfy the following perturbation
equations:%
\begin{eqnarray}
T_{1}^{1}+p &=&0  \nonumber \\
T_{2}^{2}+p &=&0  \nonumber \\
\left( T_{3}^{3}+p\right) \left( T_{4}^{4}+p\right) -T_{4}^{3}T_{3}^{4} &=&0
\nonumber \\
T_{3}^{1} &=&0\qquad \qquad T_{1}^{3}\neq 0  \nonumber \\
T_{1}^{2} &=&0=T_{2}^{1}  \nonumber \\
T_{4}^{2} &=&0=T_{2}^{4}  \label{ebo28a}
\end{eqnarray}

The first or second of these equations may be used to calculate the internal
pressure $p$ whilst the density $\rho $ is calculated using%
\begin{equation}
\rho =T_{a}^{a}+3p  \label{ebo29}
\end{equation}%
where the repeated index indicates summation. It is perhaps worthy of note
that although the velocity four vector $u_{RW}^{1}=0$ for the standard
Robertson-Walker case, the system of equations (\ref{ebo28a}) do not imply
that this condition is retained for the rotating source.

The angular velocity of the source will be denoted by $L\left( \xi ,\eta
\right) $ where:%
\begin{equation}
L\left( \xi ,\eta \right) \equiv \frac{u^{3}}{u^{4}}=\frac{T_{4}^{3}}{%
T_{4}^{4}+p}  \label{ebo30}
\end{equation}%
Since the fourth of conditions (\ref{ebo28a}) $T_{3}^{1}=0$ may be solved
immediately to give:%
\begin{equation}
Y_{\eta }=X_{\xi }+\frac{h\left( \xi \right) }{(1-k\xi ^{2})^{\frac{1}{2}%
}\xi ^{4}R^{3}}  \label{ebo21ab}
\end{equation}%
where $h\left( \xi \right) $ is an arbitrary function of $\xi $, it follows
from (\ref{ebo30}) that the angular velocity of the source is given by: 
\begin{equation}
L\left( \xi ,\eta \right) =-\frac{q\sqrt{1-k\xi ^{2}}h_{\xi }}{16\pi \xi
^{4}R^{5}\left( \rho _{RW}+p_{RW}\right) }-qX  \label{cf57y}
\end{equation}%
where the suffix denotes a partial derivative. The density $\rho _{RW}$ and
pressure $p_{RW}$ for the standard \ Robertson-Walker metric (\ref{ebo5})
are such that:%
\begin{equation}
8\pi \left( p_{RW}+\rho _{RW}\right) =-\frac{2R_{\eta \eta }}{R}+\frac{%
2R_{\eta }^{2}}{R^{2}}+\frac{2k}{R^{2}}  \label{p_plus_rho}
\end{equation}

Moreover, a particle moving in \ the field of (\ref{ebo19s}) will have zero
angular momentum whenever $u_{3}=0$, so that the quantity: 
\begin{equation}
\frac{u_{3}}{u_{4}}=\frac{q\sin ^{2}\theta \sqrt{1-k\xi ^{2}}h_{\xi }}{16\pi
\xi ^{2}R^{3}\left( \rho _{RW}+p_{RW}\right) }  \label{ebo98k}
\end{equation}%
will also be zero for such a particle. It follows that the induced angular
velocity $\Omega \left( \xi ,\eta \right) $ of the inertial frame is given
by: 
\begin{equation}
\Omega _{f}\left( \xi ,\eta \right) =-qX\quad \text{,}  \label{ebo98s}
\end{equation}%
and that the angular velocity of a particle moving in \ the field of (\ref%
{ebo19s}) is:

\begin{equation}
\Omega _{p}\left( \xi ,\eta \right) =-\frac{q\sqrt{1-k\xi ^{2}}h_{\xi }}{%
16\pi \xi ^{4}R^{5}\left( \rho _{RW}+p_{RW}\right) }  \label{ebo99s}
\end{equation}

Clearly therefore it is the nature of $h\left( \xi \right) $ which
determines the actual angular velocity of the system and that $h=0$ defines
a non-rotating source and that the solution of (\ref{ebo21ab}) is then:

\begin{equation}
X=\Phi _{\eta }\qquad Y=\Phi _{\xi }  \label{suz1}
\end{equation}%
for some $\Phi =\Phi \left( \xi ,\eta \right) .$

\section{Simplification of the Perturbation equations}

Using the second equation $p=-T_{2}^{2}$ the first and third of (\ref{ebo28a}%
)\ \ with (\ref{ebo60p}) become:%
\begin{equation}
G_{1}^{1}-G_{2}^{2}=0  \label{ebo36b}
\end{equation}%
\begin{equation}
\left( G_{3}^{3}-G_{2}^{2}\right) \left( G_{4}^{4}-G_{2}^{2}\right)
-G_{4}^{3}G_{3}^{4}=0  \label{ebo37}
\end{equation}%
In addition the fifth and sixth of (\ref{ebo28a})\ namely, $%
T_{1}^{2}=0=T_{2}^{1}$ and $T_{4}^{2}=0=T_{2}^{4}$ will be satisfied by
taking:%
\begin{equation}
G_{1}^{2}=0  \label{ebo39k}
\end{equation}%
\begin{equation}
G_{4}^{2}=0  \label{ebo40a}
\end{equation}%
These equations may be simplified somewhat by firstly, defining a new time
variable $\tau \left( \eta \right) $ through:%
\begin{equation}
\tau \left( \eta \right) =\int \frac{d\eta }{R^{3}}  \label{ebo41}
\end{equation}%
and, secondly by introducing the functions $S\left( \tau \right) $ and $%
T\left( \tau \right) $\ \ expressed in terms of the density $\rho _{RW}$ and
pressure $p_{RW}$ for the standard \ Robertson-Walker metric as follows:%
\begin{equation}
S\left( \tau \right) \equiv \frac{1}{R^{6}T}\equiv 8\pi \left( p_{RW}+\rho
_{RW}\right) =-\frac{2R_{\tau \tau }}{R^{7}}+\frac{8R_{\tau }^{2}}{R^{8}}+%
\frac{2k}{R^{2}}  \label{ebo49}
\end{equation}%
Thirdly the equations (\ref{ebo36b}) to (\ref{ebo39k}) may be rendered
independent of $\tilde{W}$ with the aid of the following substitutions:%
\begin{eqnarray}
\tilde{J} &=&XY\sin ^{2}\theta +\frac{J}{R^{3}\xi \sqrt{1-k\xi ^{2}}} 
\nonumber \\
\tilde{U} &=&\xi ^{2}\left( 1-k\xi ^{2}\right) Y^{2}\sin ^{2}\theta +U+W 
\nonumber \\
\tilde{V} &=&V+W  \nonumber \\
\tilde{W} &=&W  \nonumber \\
\tilde{Q} &=&-\xi ^{2}R^{2}X^{2}\sin ^{2}\theta +Q-W  \label{ebo74}
\end{eqnarray}%
where each of $U$, $V$, $W$, $Q$ and $J$ are again functions of $\xi $, $%
\theta $ and $\tau $.

It follows from the transformations (\ref{ebo74}) and (\ref{ebo41}) that the
metric (\ref{ebo19s}) may now be written in the form:%
\begin{eqnarray}
d\sigma ^{2} &=&\xi ^{2}R^{2}\left\{ \frac{\left( 1+\left( Q-W\right)
q^{2}\right) R^{4}}{\xi ^{2}}d\tau ^{2}-\frac{\left( 1+\left( U+W\right)
q^{2}\right) }{\xi ^{2}\left( 1-k\xi ^{2}\right) }d\xi ^{2}\right.  \nonumber
\\
&&-\frac{2Jq^{2}}{\xi \sqrt{1-k\xi ^{2}}}d\xi d\tau -\left( 1+\left(
V+W\right) q^{2}\right) d\theta ^{2}  \nonumber \\
&&\left. -Wq^{2}\sin ^{2}\left( \theta \right) d\phi ^{2}-\sin ^{2}\left(
\theta \right) \left( d\phi +qYd\xi +qR^{3}Xd\tau \right) ^{2}\right\}
\label{new_metric}
\end{eqnarray}%
Note that the entity $d\phi +qYd\xi +qR^{3}Xd\tau $ is itself an exact
differential only in the non-rotating case when equation (\ref{suz1})
applies.

In this way the first of equations (\ref{ebo36b}) becomes: 
\begin{eqnarray}
&&\left\{ -\xi \left( 1-k\xi ^{2}\right) \left( U_{\xi }+V_{\xi }\right) -%
\frac{\xi ^{2}\left( U_{\tau \tau }-V_{\tau \tau }\right) }{R^{4}}+2\left(
U-V\right) \right.  \nonumber \\
&&\left. +\frac{2\xi ^{3}\sqrt{1-k\xi ^{2}}J_{\tau \xi }}{R^{4}}+\xi
^{2}\left( 1-k\xi ^{2}\right) Q_{\xi \xi }-\xi Q_{\xi }-Q_{\theta \theta
}\right\}  \nonumber \\
&&+\frac{\left( V_{\theta }+U_{\theta }\right) \cos \theta }{\sin \theta }-%
\frac{h^{2}\sin ^{2}\theta }{\xi ^{4}R^{4}}=0  \label{eq_1122}
\end{eqnarray}%
In addition the second of equations (\ref{ebo37}) has the form:

\begin{eqnarray}
&&\xi ^{2}\left( 1-k\xi ^{2}\right) V_{\xi \xi }-3k\xi ^{3}V_{\xi }+2\xi
V_{\xi }-\frac{\xi ^{2}V_{gg}}{R^{4}}+U_{\theta \theta }+Q_{\theta \theta } 
\nonumber \\
&&-\frac{\left( U_{\theta }+Q_{\theta }\right) \cos \theta }{\sin \theta }%
-\sin ^{2}\theta \left( \frac{h_{\xi }^{2}\left( 1-k\xi ^{2}\right) T}{2\xi
^{4}}+\frac{h^{2}}{\xi ^{4}R^{4}}\right) =0  \label{eq_43}
\end{eqnarray}%
The condition (\ref{ebo39k}) that $G_{1}^{2}=0$ becomes:%
\begin{equation}
-\frac{U_{\theta }+Q_{\theta }}{\xi }+\frac{\xi J_{\tau \theta }}{R^{4}\sqrt{%
1-k\xi ^{2}}}+Q_{\theta \xi }-\frac{V_{\xi }\cos \theta }{\sin \theta }=0
\label{eq12}
\end{equation}%
Only the remaining equation $\ $(\ref{ebo40a}) explicitly contains $W$ as
follows:%
\begin{equation}
\frac{2R_{\tau }\left( W_{\theta }-Q_{\theta }\right) }{R}+2W_{\tau \theta
}+U_{\tau \theta }-\sqrt{1-k\xi ^{2}}\left( \xi J_{\xi \theta }+J_{\theta
}\right) -\frac{V_{\tau }\cos \theta }{\sin \theta }=0  \label{eq42}
\end{equation}%
Equations (\ref{eq_1122}) to \ref{eq42}) are the final forms of the
perturbation equations which determine $U$, $V$, $W$, $Q$ and $J$ for given $%
h\left( \xi \right) $ for the metric (\ref{new_metric}).

The internal supporting pressure, calculated using the second of (\ref%
{ebo28a}) is:%
\begin{eqnarray}
8\pi p &=&q^{2}\left\{ \frac{\sqrt{1-k\xi ^{2}}}{R^{6}}\left( \xi J_{\tau
\xi }+2J_{\tau }\right) +\left( W-Q\right) \left( \frac{1}{R^{6}T}-\frac{%
3R_{\tau }^{2}}{R^{2}}\right) \right.  \nonumber \\
&&-\frac{\left( W_{\tau }-Q_{\tau }\right) R_{\tau }}{R^{7}}+\frac{k\left(
U+2Q-W\right) }{R^{2}}-\frac{\left( 2W_{\tau \tau }+U_{\tau \tau }\right) }{%
2R^{6}}  \nonumber \\
&&\left. +\frac{\left( 1-k\xi ^{2}\right) }{2R^{2}}\left( \frac{2Q_{\xi
}-U_{\xi }}{\xi }+Q_{\xi \xi }\right) -\frac{Q_{\xi }}{2\xi R^{2}}\right\} 
\nonumber \\
&&+\frac{q^{2}\left( Q_{\theta }+U_{\theta }\right) \cos \theta }{2\xi
^{2}R^{2}\sin \theta }-\frac{h^{2}q^{2}\sin ^{2}\theta }{4\xi ^{6}R^{6}}%
+8\pi p_{RW}  \label{pressurea}
\end{eqnarray}%
whilst the internal density, calculated using (\ref{ebo29}) is:

\begin{eqnarray}
8\pi \rho &=&q^{2}\left\{ \frac{\left( 1-k\xi ^{2}\right) }{R^{2}}\left( 
\frac{Q_{\xi \xi }}{2}-W_{\xi \xi }-V_{\xi \xi }+\frac{U_{\xi }}{2\xi }%
\right) -\frac{\left( W_{\theta \theta }+U_{\theta \theta }+Q_{\theta \theta
}\right) }{\xi ^{2}R^{2}}\right.  \nonumber \\
&&\frac{\left( V_{\tau \tau }-\frac{U_{\tau \tau }}{2}\right) }{R^{6}}+\frac{%
R_{\tau }}{R^{7}}\left( 3W_{\tau }+V_{\tau }+U_{\tau }\right) +\frac{%
3R_{\tau }^{2}\left( W-Q\right) }{R^{8}}-\frac{3k\left( U+W\right) }{R^{2}} 
\nonumber \\
&&+\frac{2\left( U-V\right) }{\xi ^{2}R^{2}}+\frac{1}{R^{2}}\left( 3k\xi
W_{\xi }-\frac{2W_{\xi }}{\xi }+4k\xi V_{\xi }-\frac{3V_{\xi }}{\xi }-\frac{%
Q_{\xi }}{2\xi }\right)  \nonumber \\
&&\left. +\frac{\sqrt{1-k\xi ^{2}}}{R^{6}}\left( -\frac{2\xi J_{\xi }R_{\tau
}}{R}-\frac{6JR_{\tau }}{R}+\xi J_{\tau \xi }\right) \right\}  \nonumber \\
&&+\frac{q^{2}\left( Q_{\theta }+U_{\theta }+2V_{\theta }-2W_{\theta
}\right) \cos \theta }{2\xi ^{2}R^{2}\sin \theta }-\frac{h^{2}q^{2}\sin
^{2}\theta }{4\xi ^{6}R^{6}}+8\pi \rho _{RW}  \label{density}
\end{eqnarray}

Further note that the application of the transformations (\ref{ebo74}) and (%
\ref{ebo41}) has enabled each of the equations (\ref{eq_1122}) to (\ref%
{density}) to be written in a form which is independent of both $X\left( \xi
,\tau \right) $ and $Y\left( \xi ,\tau \right) $ but explicitly contains
terms in $h\left( \xi \right) .$This is expected from (\ref{ebo98s}) since
one would not expect internal pressure and density to be dependent on the
frame dragging term $X$ \ but rather on that function defined in (\ref{cf57y}%
) and (\ref{ebo98k}) which directly determines the true angular velocity of
the source, namely $h.$ It follows that further analysis may continue
without further detailed specification of $X$ and $Y$, only $h$ needs
consideration.

Finally, the two equations (\ref{eq_1122}) to (\ref{eq12}) contain the four
unknowns $J$,$Q$, $U$, $V$ and may be used to determine families of rotating
extended sources for a range gauges and or physical conditions. As a
particular example of a solution procedure it should be noted that in cases
when $V$ is known explicitly then equation (\ref{eq_43}) may be integrated
directly to determine $U+Q$. Thus if either $U$ or $Q$ is known then the
equation (\ref{eq12}) may be integrated immediately to determine $J$.

Finally, the velocity four-vector component $u^{1}/u^{4}$ may be calculated
through 
\begin{equation}
\frac{u^{1}}{u^{4}}=\frac{T_{4}^{1}}{T_{4}^{4}+p}  \label{w56}
\end{equation}%
and so up to and including quadratic terms in $q^{2}$ it follows that:%
\begin{equation}
\frac{u^{1}}{u^{4}}=-\frac{G_{4}^{1}}{8\pi \left( \rho _{RW}+p_{RW}\right) }
\label{w57}
\end{equation}%
where $G_{4}^{1}$ is given by:%
\begin{eqnarray}
-\frac{R^{5}G_{4}^{1}}{\sqrt{1-k\xi ^{2}}} &=&\left\{ \left( W_{\tau \xi }+%
\frac{V_{\tau \xi }}{2}\right) +\frac{1}{\xi }\left( \frac{V_{\tau }}{2}%
-U_{\tau }\right) +\frac{R_{\tau }}{R}\left( W_{\xi }-Q_{\xi }\right)
\right\} \sqrt{1-k\xi ^{2}}  \nonumber \\
&&+\frac{J_{\theta \theta }}{2\xi }-\frac{\xi J}{R^{4}T}+2k\xi J+\frac{%
J_{\theta }\cos \theta }{2\xi \sin \theta }  \label{eq_41}
\end{eqnarray}%
Note also that in a similar way $u_{1}/u_{4}$ may be found using;%
\begin{equation}
\frac{u_{1}}{u_{4}}=\frac{T_{1}^{4}}{T_{4}^{4}+p}=-\frac{G_{1}^{4}}{8\pi
\left( \rho _{RW}+p_{RW}\right) }  \label{eq69}
\end{equation}%
where $G_{1}^{4}$ is

\begin{eqnarray}
&&G_{1}^{4}=-\frac{h_{\xi }q^{2}\,\sin ^{2}\theta \,\,Y\sqrt{1-k\xi ^{2}}}{%
2\,\xi ^{2}\,R^{3}}-q^{2}\left\{ -\frac{2W_{\tau \xi }+V_{\tau \xi }}{2R^{3}}%
+\frac{2U_{\tau }-V_{\tau }}{2\xi \,R^{3}}\right.  \nonumber \\
&&\left. +\frac{R_{\tau }\left( Q_{\xi }-W_{\xi }\right) }{R^{4}}+\frac{2J%
\sqrt{1-k\xi ^{2}}}{\xi \,R^{3}}-\left( \frac{J_{\theta \theta }+4J}{2\,\xi 
\sqrt{1-k\xi ^{2}}\,R^{3}}\right) \right\}  \nonumber \\
&&+\frac{q^{2}\,\cos \theta \,J_{\theta }}{2\,\sin \theta \,\xi \,\sqrt{%
1-k\xi ^{2}}\,R^{3}}  \label{eq14a}
\end{eqnarray}
Thus only (\ref{eq14a}) depends explicitly on $Y\left( \xi ,\tau \right) $
and is a further manifestation of the frame dragging effect expressed in (%
\ref{ebo21ab}). This may be removed by choosing $Y=0$.

\section{Characterisation of angular velocity leading to analytic solutions}

Consider first equations the three equations (\ref{eq_1122}) to (\ref{eq12})
and note that particular solutions may be found in principle by setting:%
\begin{equation}
U\left( \xi ,\theta ,\tau \right) =\frac{u_{1}}{\xi ^{6}}\sin ^{2}\theta
+u_{2}  \label{puaa}
\end{equation}%
\begin{equation}
V\left( \xi ,\theta ,\tau \right) =\frac{u_{3}\sin ^{2}\theta }{\xi ^{6}}%
+u_{4}  \label{pvaa}
\end{equation}%
\begin{equation}
J\left( \xi ,\theta ,\tau \right) =\sqrt{1-k\xi ^{2}}\left( u_{5}\sin
^{2}\theta +u_{6}\right)  \label{pjaaa}
\end{equation}%
\begin{equation}
Q\left( \xi ,\theta ,\tau \right) =u_{7}\sin ^{2}\theta +u_{8}  \label{pjaa}
\end{equation}%
where $u_{i}$ are functions of $\xi $ and $\tau $ alone. It is straight
forward to show that equation (\ref{eq_43}) may be used to determine $u_{7}$
and $u_{4}$ since:%
\begin{eqnarray}
\xi ^{6}u_{7} &=&-\frac{\xi ^{2}\left( h^{2}+u_{3_{\tau \tau }}\right) }{%
2R^{4}}+\frac{\xi ^{2}(1-k\xi ^{2})}{2}\left( u_{3_{\xi \xi }}-\frac{h_{\xi
}^{2}T}{2}\right)  \nonumber \\
&&+\frac{9k\xi ^{3}u_{3_{\xi }}}{2}-12k\xi ^{2}u_{3}-5\xi u_{3_{\xi
}}+15u_{3}-u_{1}  \label{pu7b}
\end{eqnarray}%
and $u_{4}=0$. Moreover direct substitution of these relationships into (\ref%
{eq12}) shows that this equation is satisfied provided that $u_{5}$ is such
that:%
\begin{equation}
\frac{\partial u_{5}}{\partial \tau }=\Psi _{1}R^{4}T+\Psi _{2}R^{4}+\Psi
_{3}  \label{pu5d}
\end{equation}%
where $\Psi _{1}$,$\Psi _{2}$ and $\Psi _{3}$ are fully defined in terms of $%
h$, $u_{1}$ and $u_{3}$ and are presented in the appendix. The equation (\ref%
{pu7b}) and (\ref{pu5d}) together with $u_{4}=0$ may then be substituted
directly into equation (\ref{eq_1122}) to produce a very lengthy
relationship, reproduced in the appendix, having the general form:%
\begin{equation}
\Psi _{4}\sin ^{2}\theta +\Psi _{5}=0  \label{pu1122}
\end{equation}%
where $\Psi _{4}$ and $\Psi _{5}$ are again fully defined in terms of $h$, $%
u_{1}$ and $u_{3}$ and their partial derivatives with respect to $\xi $ and $%
\tau $.

Inspection shows that the equation $\Psi _{4}=0$ may in general terms only
be solved numerically for $u_{1}$(say) in terms any given $u_{3}\left( \xi
,\tau \right) $ and $h\left( \xi \right) $. Remarkably, however there is at
least one case where considerable simplification is possible, namely when:%
\begin{equation}
h\left( \xi \right) =n\xi ^{5}  \label{pha}
\end{equation}%
with constant $n$, for which a basic separation of variables approach
reveals that:%
\begin{equation}
u_{1}\left( \xi ,\tau \right) =n^{2}\xi ^{10}\left( \psi _{1}+\psi _{2}\xi
^{2}\right) \qquad \qquad u_{3}\left( \xi ,\tau \right) =n^{2}\xi ^{10}\psi
_{3}  \label{p13a}
\end{equation}%
where $\psi _{1}$, $\psi _{2}$ and $\psi _{3}$ are functions of $\tau $
alone which satisfy:%
\begin{equation}
125k^{2}T-\frac{10k\left( 1+\psi _{3_{\tau \tau }}\right) +\psi _{2_{\tau
\tau }}}{R^{4}}-24k\left( \psi _{2}+10k\psi _{3}\right) =0  \label{p1122b}
\end{equation}%
\begin{equation}
-\frac{225kT}{2}+\frac{7\psi _{3_{\tau \tau }}-\psi _{1_{\tau \tau }}+5}{%
R^{4}}+4k\left( 49\psi _{3}-2\psi _{1}\right) +14\psi _{2}=0  \label{p1122c}
\end{equation}%
Although these equations need to be solved numerically when $k=1$, $-1$ the
case $k=0$ yields the analytic result that:

\begin{equation}
7\psi _{3_{\tau \tau }}-\psi _{1_{\tau \tau }}+5=0\qquad \psi _{2}=0
\label{p1122d}
\end{equation}%
In the following the family of solutions presented will be based upon
equation (\ref{p1122d}). Thus using (\ref{pha}) and (\ref{ebo41})
substituted in (\ref{ebo99s}) and the angular velocity $\Omega _{p}\left(
\xi ,\tau \right) $ of a particle in the field of (\ref{new_metric}) is:%
\begin{equation}
\Omega _{p}\left( \xi ,\tau \right) =-\frac{nq}{16\pi R^{2}\left( \rho
_{RW}+p_{RW}\right) }  \label{pont5}
\end{equation}%
It follows that the angular velocity is purely time dependent. The frame
dragging effect (\ref{ebo98s}) is 
\begin{equation}
\Omega _{f}=-qR^{3}X  \label{pont7}
\end{equation}%
where from (\ref{ebo21ab}) and (\ref{ebo41}) $X\left( \xi ,\tau \right) $ is
determined through:%
\begin{equation}
R^{3}X_{\xi }=Y_{\tau }-n\xi  \label{pont10}
\end{equation}%
In cases when $Y=0$ so that the frame dragging effect due to rotation is
removed from (\ref{eq14a}) then (\ref{pont5}) becomes:%
\begin{equation}
\Omega _{f}=\frac{nq\xi ^{2}}{2}  \label{pont15}
\end{equation}%
where for choice $X\left( 0,\tau \right) =0$.

\section{Development of solutions with k=0 and with purely time dependent
angular velocity}

Thus using the solution procedure outlined in the previous section with (\ref%
{p1122d}) it is straight forward to show that the corresponding solutions of
(\ref{eq_1122}) to (\ref{eq42}) are:

\begin{equation}
U\left( \xi ,\theta ,\tau \right) =n^{2}\psi _{1}\xi ^{4}\sin ^{2}\theta
+n^{2}z_{2}  \label{qub}
\end{equation}

\begin{equation}
V\left( \xi ,\theta ,\tau \right) =\frac{n^{2}}{7}\left( \psi _{1}-\frac{%
5\tau ^{2}}{2}\right) \xi ^{4}\sin ^{2}\theta  \label{pvb}
\end{equation}

\begin{equation}
J\left( \xi ,\theta ,\tau \right) =n^{2}z_{5}\sin ^{2}\theta +n^{2}z_{6}
\label{pjb}
\end{equation}

\begin{equation}
Q\left( \xi ,\theta ,\tau \right) =n^{2}\left\{ \xi ^{4}\left( \frac{3\psi
_{1}}{7}-\frac{25T}{4}-\frac{25\tau ^{2}}{7}\right) -\frac{\xi ^{6}\left(
\psi _{1_{\tau \tau }}+2\right) }{14R^{4}}\right\} \sin ^{2}\theta
+n^{2}z_{8}  \label{pqb}
\end{equation}

\begin{equation}
W\left( \xi ,\theta ,\tau \right) =\frac{n^{2}z_{9}\sin ^{2}\theta }{R}+%
\frac{n^{2}z_{10}}{R}  \label{pwb}
\end{equation}%
where $z_{5}\left( \xi ,\tau \right) $, $z_{6}\left( \xi ,\tau \right) $ and 
$z_{9}\left( \xi ,\tau \right) $ are determined in terms of $\psi _{1}\left(
\tau \right) $, $z_{2}\left( \xi ,\tau \right) $ and $z_{8}\left( \xi ,\tau
\right) $ through:%
\begin{equation}
\frac{\partial z_{5}}{\partial \tau }=\xi ^{2}R^{4}\left( \frac{75T}{4}%
+10\tau ^{2}\right) +\frac{5\xi ^{4}}{14}\left( \psi _{1_{\tau \tau
}}+2\right)  \label{z5dg}
\end{equation}%
\begin{eqnarray}
\frac{\partial ^{2}z_{6}}{\partial \xi \partial \tau } &=&-R^{4}\left\{ \xi
\left( \frac{25T}{4}+\frac{45\tau ^{2}}{14}+\frac{5\psi _{1}}{7}\right) +%
\frac{z_{8_{\xi \xi }}}{2\xi }-\frac{z_{8_{\xi }}+z_{2_{\xi }}}{2\xi ^{2}}+%
\frac{z_{2}}{\xi ^{2}}\right\}  \nonumber \\
&&-\frac{\xi ^{3}}{14}\left( \psi _{1_{\tau \tau }}+2\right) +\frac{%
z_{2_{\tau \tau }}}{2\xi }  \label{z6dgdx}
\end{eqnarray}%
\begin{eqnarray}
\frac{\partial z_{9}}{\partial \tau } &=&\xi ^{4}R_{\tau }\left( \frac{3\psi
_{1}}{7}-\frac{25T}{4}-\frac{25\tau ^{2}}{7}\right) -\frac{\xi ^{6}R_{\tau }%
}{14R^{4}}\left( \psi _{1_{\tau \tau }}+2\right)  \nonumber \\
&&-\frac{\xi ^{4}R}{28}\left( 5\tau +13\psi _{1_{\tau }}\right) +\frac{R}{2}%
\left( \xi z_{5_{\xi }}+z_{5}\right)  \label{z9dg}
\end{eqnarray}%
The function $z_{10}\left( \xi ,\tau \right) $ is arbitrary. The supporting
internal pressure (\ref{pressurea}) is given by:

\begin{eqnarray}
8\pi p &=&8\pi p_{RW}+n^{2}q^{2}\sin ^{2}\theta \left\{ \frac{25\xi ^{2}T}{%
8R^{2}}+\frac{\xi ^{4}}{14R^{6}T}\left( 25\tau ^{2}-3\psi _{1}\right) +\frac{%
z_{9}}{2R^{7}T}\right.  \nonumber \\
&&\left. +\frac{\xi ^{6}}{28R^{10}T}\left( \psi _{1_{\tau \tau }}+2\right) +%
\frac{25\xi ^{4}}{8R^{6}}\right\}  \nonumber \\
&&+n^{2}q^{2}\xi ^{2}\left\{ -\frac{25T}{2R^{2}}-\frac{95\tau ^{2}}{14R^{2}}+%
\frac{5\psi _{1}}{7R^{2}}\right\} -\frac{\xi ^{4}}{7R^{6}}\left( \psi
_{1_{\tau \tau }}+2\right)  \nonumber \\
&&+n^{2}q^{2}\left\{ -\frac{z_{8}}{R^{6}T}+\frac{z_{9}}{2R^{7}T}+\frac{%
3z_{8}R_{\tau }^{2}}{R^{8}}+\frac{z_{8_{\tau }}R_{\tau }}{R^{7}}+\frac{%
z_{10_{\tau }}R_{\tau }}{R^{8}}\right.  \nonumber \\
&&\left. +\frac{z_{8_{\xi }}}{\xi R^{2}}-\frac{z_{2}}{\xi ^{2}R^{2}}+\frac{%
2z_{6_{\tau }}}{R^{6}}-\frac{z_{10_{\tau \tau }}}{R^{7}}\right\}
\label{pres}
\end{eqnarray}%
with:%
\begin{equation}
8\pi p_{RW}=\frac{1}{R^{6}T}-\frac{3R_{\tau }^{2}}{R^{8}}  \label{RW_pres}
\end{equation}%
and the internal density is (\ref{density}) is%
\begin{eqnarray}
8\pi \rho &=&8\pi \rho _{RW}+n^{2}q^{2}\sin ^{2}\theta \left\{ \frac{\xi ^{2}%
}{R^{2}}\left( -\frac{25T}{4}+5\tau ^{2}+6\psi _{1}\right) -\frac{\xi
^{4}R_{\tau }}{4R^{7}}\left( 5\tau +\psi _{1_{\tau }}\right) \right. 
\nonumber \\
&&\left. -\frac{R_{\tau }}{2R^{7}}\left( \xi z_{5_{\xi }}+9z_{5}\right) +%
\frac{1}{R^{3}}\left( -\frac{2z_{9_{\xi }}}{\xi }+\frac{6z_{9}}{\xi ^{2}}%
-z_{9_{\xi \xi }}\right) -\frac{\xi ^{4}}{4R^{6}}\right\}  \nonumber \\
&&+n^{2}q^{2}\left\{ -\frac{\xi ^{2}}{7R^{2}}\left( \frac{5\tau ^{2}}{2}%
+13\psi _{1}\right) +\frac{1}{\xi R^{2}}\left( z_{2_{\xi }}+\frac{z_{2}}{\xi 
}\right) \right.  \nonumber \\
&&-\frac{1}{R^{3}}\left( \frac{2z_{10_{\xi }}}{\xi }+\frac{4z_{9}}{\xi ^{2}}%
+z_{10_{\xi \xi }}\right)  \nonumber \\
&&\left. -\frac{3z_{8}R_{\tau }^{2}}{R^{8}}+\frac{R_{\tau }}{R^{7}}\left(
-2\xi z_{6_{\xi }}+z_{2_{\tau }}-6z_{6}+\frac{3z_{10_{\tau }}}{R}\right)
\right\}  \label{genden}
\end{eqnarray}%
with:%
\begin{equation}
8\pi \rho _{RW}=\frac{3R_{\tau }^{2}}{R^{8}}  \label{RW_den}
\end{equation}%
Also using (\ref{w57}) and (\ref{eq_41}) it may be shown that: 
\begin{eqnarray}
\frac{1}{RT}\frac{u^{1}}{u^{4}} &=&n^{2}q^{2}\left\{ -\frac{\xi z_{6}}{R^{4}T%
}+\frac{z_{10_{\tau \xi }}-z_{8_{\xi }}R_{\tau }}{R}+\frac{2z_{5}-z_{2_{\tau
}}}{\xi }\right\}  \nonumber \\
&&+n^{2}q^{2}\sin ^{2}\theta \left\{ \frac{\xi }{2}\left( z_{5_{\xi \xi }}-%
\frac{2z_{5}}{R^{4}T}\right) +z_{5_{\xi }}-\frac{3z_{5}}{\xi }-\frac{5\xi
^{3}\left( \tau +\psi _{1_{\tau }}\right) }{2}\right\}  \label{ein41aa}
\end{eqnarray}%
Inspection of (\ref{ein41aa}) shows that in general terms $u^{1}\left( \xi
,\theta ,\tau \right) $ although it is possible to choose $z_{10}$ so that $%
u^{1}\left( \xi ,0,\tau \right) =0$. Thus $u^{1}=0$ is possible for a
particle moving in the field of (\ref{new_metric}) only for an observer
situated on the axis of rotation. In this case:%
\begin{equation}
\frac{\partial ^{2}z_{10}}{\partial \tau \partial \xi }=\frac{z_{6}\xi }{%
R^{3}T}+z_{8_{\xi }}R_{\tau }-\frac{R}{\xi }\left( 2z_{5}-z_{2_{\tau
}}\right)  \label{z10dg}
\end{equation}

\section{A subclass of solutions}

Consider now the particular case:%
\begin{equation}
\psi _{1}=-\tau ^{2}\qquad z_{2}=\omega _{2}\xi ^{2}\qquad z_{8}=\omega
_{8}\xi ^{2}  \label{example1}
\end{equation}%
where $\omega _{2}$ and $\omega _{8}$ are functions of $\tau $. The
equations (\ref{z5dg}) to (\ref{z9dg}) become:%
\begin{equation}
z_{5}=\psi _{5}\xi ^{2}\qquad \qquad \frac{d\psi _{5}}{d\tau }=R^{4}\left( 
\frac{75T}{4}+10\tau ^{2}\right)  \label{z5_ex}
\end{equation}%
\begin{equation}
z_{6}=\psi _{6}\xi ^{2}\qquad \qquad \frac{d\psi _{6}}{d\tau }=-R^{4}\left( 
\frac{25T}{8}+\frac{5\tau ^{2}}{4}\right) +\frac{\omega _{2_{\tau \tau }}}{4}
\label{z6_ex}
\end{equation}%
Moreover%
\begin{equation}
z_{9}=\alpha _{2}\xi ^{2}+\alpha _{4}\xi ^{4}  \label{z9_ex}
\end{equation}%
\begin{equation}
\frac{d\alpha _{2}}{d\tau }=\frac{3\psi _{5}R}{2}\qquad \qquad \frac{d\alpha
_{4}}{d\tau }=\frac{3\tau R}{4}-R_{\tau }\left( \frac{25T}{4}+4\tau
^{2}\right)  \label{z9Ex1}
\end{equation}%
where $\psi _{5}$, $\psi _{6}$, $\alpha _{2}$ and $\alpha _{4}$ are
functions of $\tau $ alone. The condition (\ref{z10dg}) is:%
\begin{equation}
z_{10}=\beta _{2}\xi ^{2}+\beta _{4}\xi ^{4}  \label{z10_ex}
\end{equation}%
\begin{equation}
\frac{d\beta _{2}}{d\tau }=\omega _{8}R_{\tau }+R\left( \frac{\omega
_{2_{\tau }}}{2}-\psi _{5}\right) \qquad \qquad \frac{d\beta _{4}}{d\tau }=%
\frac{\psi _{6}}{4R^{3}T}  \label{z10_ex1}
\end{equation}%
where $\beta _{2}=\beta _{2}\left( \tau \right) $ and $\beta _{4}=\beta
_{4}\left( \tau \right) $.

The supporting internal pressure is given by:%
\begin{eqnarray}
8\pi p &=&8\pi p_{RW}+n^{2}q^{2}\sin ^{2}\theta \left\{ \xi ^{2}\left( \frac{%
25T}{8R^{2}}+\frac{\alpha _{2}}{2R^{7}T}\right) \right.  \nonumber \\
&&\left. +\xi ^{4}\left( \frac{25}{8R^{6}}+\frac{2\tau ^{2}}{R^{6}T}+\frac{%
\alpha _{4}}{2R^{7}T}\right) \right\}  \nonumber \\
&&+n^{2}q^{2}\xi ^{2}\left\{ -\frac{75T}{4R^{2}}-\frac{\omega _{8}}{R^{6}T}+%
\frac{\beta _{2}}{2R^{7}T}+\frac{3\omega _{8}R_{\tau }^{2}}{R^{8}}+\frac{%
R_{\tau }}{R^{7}}\left( \omega _{8_{\tau }}+\frac{\beta _{2_{\tau }}}{R}%
\right) \right.  \nonumber \\
&&\left. -\frac{10\tau ^{2}}{R^{2}}+\frac{\omega _{2_{\tau \tau }}}{2R^{6}}-%
\frac{\beta _{2_{\tau \tau }}}{R^{7}}\right\}  \nonumber \\
&&+n^{2}q^{2}\xi ^{4}\left\{ \frac{\beta _{4}}{2R^{7}T}+\frac{\beta
_{4_{\tau }}R_{\tau }}{R^{8}}-\frac{\beta _{4}}{R^{7}}\right\} +\frac{%
n^{2}q^{2}}{R^{2}}\left\{ 2\omega _{8}-\omega _{2}\right\}  \label{ppresa}
\end{eqnarray}%
whilst the density is:%
\begin{eqnarray}
8\pi \rho &=&8\pi \rho _{RW}+n^{2}q^{2}\sin ^{2}\theta \left\{ \xi
^{2}\left( -\frac{25T}{4R^{2}}-\frac{11\psi _{5}R_{\tau }}{2R^{7}}-\frac{%
\tau ^{2}}{R^{2}}-\frac{14\alpha _{4}}{R^{3}}\right) \right.  \nonumber \\
&&\left. -\frac{\xi ^{4}}{4R^{6}}\left( 1+\frac{3\tau R_{\tau }}{R}\right)
\right.  \nonumber \\
&&+n^{2}q^{2}\xi ^{2}\left\{ -\frac{3\omega _{8}R_{\tau }^{2}}{R^{8}}+\frac{%
\left( \omega _{2_{\tau }}-10\psi _{6}\right) R_{\tau }}{R^{7}}+\frac{3\tau
^{2}}{2R^{2}}-\frac{4\alpha _{4}}{R^{3}}\right\}  \nonumber \\
&&+n^{2}q^{2}\left\{ \frac{3\omega _{2}}{R^{2}}-\frac{4\alpha _{2}}{R^{3}}%
\right\}  \label{pdena}
\end{eqnarray}

Notice that in region of $\xi =0$ that both the expressions for pressure and
density are well behaved and in particular when $\xi =0$ the pressure and
density of the centre of the rotating source are 
\begin{equation}
8\pi p\left( 0,\tau \right) =8\pi p_{RW}+\frac{n^{2}q^{2}}{R^{2}}\left\{
2\omega _{8}-\omega _{2}\right\}  \label{p0}
\end{equation}%
\begin{equation}
8\pi \rho \left( 0,\tau \right) =8\pi \rho _{RW}++n^{2}q^{2}\left\{ \frac{%
3\omega _{2}}{R^{2}}-\frac{4\alpha _{2}}{R^{3}}\right\}  \label{rho0}
\end{equation}%
Note that the centre of the source will be non-rotating $8\pi p\left( 0,\tau
\right) =8\pi p_{RW}$ and $8\pi \rho \left( 0,\tau \right) =8\pi \rho _{RW}$
when:%
\begin{equation}
\omega _{2}=\frac{4\alpha _{2}}{3R}\qquad \qquad \omega _{8}=\frac{2\alpha
_{2}}{3R}  \label{omrel}
\end{equation}%
For this solution (\ref{ein41aa}) defining $u^{1}$ becomes:%
\begin{equation}
\frac{u^{1}}{u^{4}}=n^{2}q^{2}\xi ^{3}RT\sin ^{2}\theta \left( \frac{5\tau }{%
2}-\frac{\psi _{5}}{R^{4}T}\right)  \label{pg41a}
\end{equation}

\section{Robertson-Walker dust endowed with rotation}

As a specific example of the above subclass consider the case of dust
endowed with rotation so that:%
\begin{equation}
R\left( \tau \right) =\kappa \tau ^{\frac{2}{3}}\qquad \qquad T\left( \tau
\right) =\frac{3\tau ^{2}}{4}  \label{dust18}
\end{equation}%
where $\kappa $ is constant \ and suppose also that $8\pi p\left( 0,\tau
\right) =8\pi p_{RW}$ and $8\pi \rho \left( 0,\tau \right) =8\pi \rho _{RW}$
in (\ref{p0}) and (\ref{rho0}), then (\ref{qub}) to (\ref{pwb}) become:%
\begin{equation}
U\left( \xi ,\theta ,\tau \right) =-\frac{207\kappa ^{4}\tau ^{\frac{20}{3}%
}n^{2}\xi ^{2}}{544}-n^{2}\tau ^{2}\xi ^{4}\sin ^{2}\theta  \label{udust}
\end{equation}%
\begin{equation}
V\left( \xi ,\theta ,\tau \right) =-\frac{n^{2}\tau ^{2}\xi ^{4}\sin
^{2}\theta }{2}  \label{vdust}
\end{equation}%
\begin{equation}
J\left( \xi ,\theta ,\tau \right) =-\frac{1155\kappa ^{4}\tau ^{\frac{17}{3}%
}n^{2}\xi ^{2}}{272}\sin ^{2}\theta  \label{jdust}
\end{equation}%
\begin{equation}
Q\left( \xi ,\theta ,\tau \right) =-\frac{139\tau ^{2}n^{2}\xi ^{4}}{16}\sin
^{2}\theta -\frac{207\kappa ^{4}\tau ^{\frac{20}{3}}n^{2}\xi ^{2}}{1088}
\label{qdust}
\end{equation}%
\begin{equation}
W\left( \xi ,\theta ,\tau \right) =\frac{n^{2}\kappa ^{4}\tau ^{6}\sin
^{2}\theta }{64}\left( \frac{945\tau ^{\frac{2}{3}}\xi ^{2}}{17}-\frac{%
121\xi ^{4}}{\kappa ^{4}\tau ^{4}}\right) -\frac{837\kappa ^{4}\tau ^{\frac{%
20}{3}}n^{2}\xi ^{2}}{1088}  \label{wdust}
\end{equation}%
and (\ref{ppresa}), (\ref{pdena}) and (\ref{pg41a}) become respectively:%
\begin{equation}
8\pi p=\frac{5n^{2}q^{2}\sin ^{2}\theta }{16\kappa ^{2}}\left\{ \frac{%
159\tau ^{\frac{2}{3}}\xi ^{2}}{17}+\frac{29\xi ^{4}}{2\tau ^{4}\kappa ^{4}}%
\right\} +\frac{7715\tau ^{\frac{2}{3}}n^{2}q^{2}\xi ^{2}}{272\kappa ^{2}}
\label{pdust}
\end{equation}%
\begin{equation}
8\pi \rho =\frac{4}{3\tau ^{6}\kappa ^{6}}+\frac{3n^{2}q^{2}}{4\kappa ^{2}}%
\sin ^{2}\theta \left\{ \frac{945\tau ^{\frac{2}{3}}\xi ^{2}}{136}-\frac{\xi
^{4}}{\tau ^{4}\kappa ^{4}}\right\} +\frac{2915\tau ^{\frac{2}{3}%
}n^{2}q^{2}\xi ^{2}}{272\kappa ^{2}}  \label{ddust}
\end{equation}%
and%
\begin{equation}
\frac{u^{1}}{u^{4}}=-\frac{645\kappa ^{6}\tau ^{7}n^{2}q^{2}\xi ^{3}\sin
^{2}\theta }{272}  \label{gdust}
\end{equation}

\section{Conclusion}

Einstein's equations for a Robertson-Walker source endowed with rotation up
to and including quadratic terms in angular velocity parameter have been
presented. It has been shown that a family of analytic solutions of the
equations are possible for the case when $k=0$ and the angular velocity of
the fluid is purely time dependent. The corresponding density and supporting
internal density are explicitly presented in a form containing perturbations
from their respective Robertson-Walker counterparts. A subclass of the
solutions merges seamlessly with the Robertson-Walker source at the origin.
The work presented here is very much a preliminary investigation and further
research is now being conducted since it is possible that further
mathematical analysis will reveal new analytic solutions of Einstein's
equations with for example spatially varying fluid angular velocity.
Moreover a numerical analysis will be applied to reveal a broader range of
properties of the perturbation equations for varying forms of $h=h\left( \xi
\right) $ and for $k=1,0,-1$. Clearly, the determination of rotating sources
incorporating graviational radiation would also be an important development.
Further work will also address gauge issues since the analytic solutions
presented in this paper have the property that the velocity four vector
components have the general property that $u^{2}=0$ and $u^{1}\neq 0$ except
at the poles where $u^{1}=0$. It would be interesting to consider solutions
for which $u^{2}=0$ and $u^{1}=0$ for all values of $\xi $, $\theta $ and $%
\tau $.

Whilst perturbation analyses are of considerable importance in the General
Theory of Relativity there is no doubt that the major goal for future
research must be in the determination of exact solutions of Einstein's
equations for rotating sources with physically realistic properties.
According to Bradley \textit{et al }\cite{Brad} there is currently an
`embarrrassing hiatus' in the availability of such solutions. It is the hope
that the perturbation analysis presented above may provide a signpost which
lead to the possible discovery of such solutions assuming that they exist.

\section{Acknowledgments}

I would like to record my sincere thanks for the helpful encouragement of
Professor Bill Bonnor at Queen Mary \& Westfield College, London during the
past year and also, Professors Leonid Grishchuk, Mike Edmunds and Peter
Blood for making me so welcome during my visit to the Physics and Astronomy
Department in Cardiff. I am also much indebted to those at the University of
Glamorgan who made the visit to Cardiff possible.

\bigskip \newpage

\appendix 

\section{Expressions terms used in section 4}

\bigskip In the following:

\begin{equation}
P=\sqrt{1-k\xi ^{2}}  \tag{A0}
\end{equation}%
so that:

\begin{equation}
\Psi _{1}\left( \xi ,\tau \right) =\frac{h_{\xi }h_{\xi \xi }\,P^{2}}{2\,\xi
^{7}}-\frac{3\,h_{\xi }^{2}\,P^{2}}{4\,\xi ^{8}}-\frac{h_{\xi }^{2}}{2\,\xi
^{8}}  \tag{A1}
\end{equation}

\begin{eqnarray}
\Psi _{2}\left( \xi ,\tau \right) &=&-\frac{\,u_{3_{\xi \xi \xi }}P^{2}}{%
2\,\xi ^{7}}+\frac{6\,u_{3_{\xi \xi }}\,P^{2}}{\xi ^{8}}-\frac{30\,u_{3_{\xi
}}\,P^{2}}{\xi ^{9}}+\frac{60\,u_{3}\,P^{2}}{\xi ^{10}}  \nonumber \\
&&+\frac{3u_{3_{\xi \xi }}\,}{2\,\xi ^{8}}-\frac{29u_{3_{\xi }}\,}{2\,\xi
^{9}}+\frac{u_{1_{\xi }}}{\xi ^{9}}+\frac{42\,u_{3}}{\xi ^{10}}-\frac{%
6\,u_{1}}{\xi ^{10}}  \TCItag{A2}
\end{eqnarray}

\begin{equation}
\Psi _{3}\left( \xi ,\tau \right) =\frac{h\,h_{\xi }}{\xi ^{7}}+\frac{%
u_{3_{\tau \tau \xi }}}{2\,\xi ^{7}}-\frac{5\,h^{2}}{2\,\xi ^{8}}-\frac{%
5u_{3_{\tau \tau }}\,}{2\,\xi ^{8}}  \tag{A3}
\end{equation}

\begin{eqnarray}
\Psi _{4}\left( \xi ,\tau \right) &=&\frac{\,h_{\xi }h_{\xi \xi \xi
}\,P^{4}\,T}{2\,\xi ^{2}}+\frac{h_{\xi \xi }^{2}\,P^{4}\,T}{2\,\xi ^{2}}-%
\frac{3h_{\xi }\,h_{\xi \xi }\,P^{4}\,T}{\xi ^{3}}+\frac{3\,h_{\xi
}^{2}\,P^{4}\,T}{\xi ^{4}}  \nonumber \\
&&-\frac{5\,h_{\xi }h_{\xi \xi }\,P^{2}\,T}{2\,\xi ^{3}}+\frac{9\,h_{\xi
}^{2}\,P^{2}\,T}{2\,\xi ^{4}}+\frac{h_{\xi }^{2}\,T}{2\,\xi ^{4}}+\frac{%
h\,h_{\xi \xi }\,P^{2}}{\xi ^{2}\,R^{4}}+\frac{h_{\xi }^{2}\,P^{2}}{\xi
^{2}\,R^{4}}  \nonumber \\
&&+\frac{\,u_{3_{\tau \tau \xi \xi }}P^{2}}{2\,\xi ^{2}\,R^{4}}-\frac{%
10\,hh_{\xi }\,P^{2}}{\xi ^{3}\,R^{4}}-\frac{5u_{3_{\tau \tau \xi }}\,P^{2}}{%
\xi ^{3}\,R^{4}}+\frac{15\,h^{2}\,P^{2}}{\xi ^{4}\,R^{4}}  \nonumber \\
&&+\frac{15\,\,u_{3_{\tau \tau }}P^{2}}{\xi ^{4}\,R^{4}}-\frac{hh_{\xi }\,}{%
\xi ^{3}\,R^{4}}-\frac{u_{3_{\tau \tau \xi }}}{2\,\xi ^{3}\,R^{4}}+\frac{%
2u_{3_{\tau \tau }}}{\xi ^{4}\,R^{4}}-\frac{u_{1_{\tau \tau }}}{\xi
^{4}\,R^{4}}-\frac{u_{3_{\xi \xi \xi \xi }}\,P^{4}}{2\,\xi ^{2}}  \nonumber
\\
&&+\frac{15u_{3_{\xi \xi \xi }}\,\,P^{4}}{2\,\xi ^{3}}-\frac{54\,u_{3_{\xi
\xi }}\,P^{4}}{\xi ^{4}}+\frac{210u_{3_{\xi }}\,\,P^{4}}{\xi ^{5}}-\frac{%
360\,u_{3}\,P^{4}}{\xi ^{6}}+\frac{3u_{3_{\xi \xi \xi }}\,P^{2}}{\xi ^{3}} 
\nonumber \\
&&-\frac{79u_{2_{\xi \xi }}\,\,P^{2}}{2\,\xi ^{4}}+\frac{u_{1_{\xi \xi
}}\,P^{2}}{\xi ^{4}}+\frac{427\,\,u_{3_{\xi }}P^{2}}{2\,\xi ^{5}}-\frac{%
13u_{1_{\xi }}\,\,P^{2}}{\xi ^{5}}-\frac{456\,u_{3}\,P^{2}}{\xi ^{6}}+ 
\nonumber \\
&&\frac{48\,u_{1}\,P^{2}}{\xi ^{6}}-\frac{3u_{3_{\xi \xi }}\,}{2\,\xi ^{4}}+%
\frac{25u_{3_{\xi }}\,}{2\,\xi ^{5}}+\frac{u_{1_{\xi }}}{\xi ^{5}}-\frac{%
34\,u_{3}}{\xi ^{6}}+\frac{2\,u_{1}}{\xi ^{6}}  \TCItag{A4}
\end{eqnarray}

\begin{eqnarray}
\Psi _{5}\left( \xi ,\tau \right) &=&\frac{h_{\xi }^{2}\,P^{2}\,T}{2\,\xi
^{4}}+\frac{2\,\,u_{6_{\tau \xi }}\xi ^{3}\,P^{2}}{R^{4}}+\frac{%
2\,u_{6_{\tau }}\,\xi ^{2}\,P^{2}}{R^{4}}-\frac{2\,u_{6_{\tau }}\,\xi ^{2}}{%
R^{4}}-\frac{u_{2_{\tau \tau }}\,\xi ^{2}}{R^{4}}  \nonumber \\
&&+\frac{h^{2}}{\xi ^{4}\,R^{4}}+\frac{u_{3_{\tau \tau }}}{\xi ^{4}\,R^{4}}%
+u_{8_{\xi \xi }}\,\xi ^{2}\,P^{2}-u_{2_{\xi }}\,\xi \,P^{2}-\frac{%
\,u_{3_{\xi \xi }}P^{2}}{\xi ^{4}}+\frac{9u_{3_{\xi }}\,P^{2}}{\xi ^{5}} 
\nonumber \\
&&-\frac{24\,u_{3}\,P^{2}}{\xi ^{6}}-u_{8_{\xi }}\xi +\frac{u_{3_{\xi }}}{%
\xi ^{5}}-\frac{4\,u_{3}}{\xi ^{6}}+\frac{4\,u_{1}}{\xi ^{6}}+2\,u_{2} 
\TCItag{A5}
\end{eqnarray}

\end{document}